# Refractive index tomography with structured illumination

Shwetadwip Chowdhury[1], Will J. Eldridge[1], Adam Wax[1], and Joseph A. Izatt[1,*]

**Authors information:**
[1]Duke University, Biomedical Engineering Department
1427 FCIEMAS, 101 Science Drive, Box 90281
Durham NC, 27708

**\*Corresponding Author:** joseph.izatt@duke.edu

## Abstract

This work introduces a novel reinterpretation of structured illumination (SI) microscopy for coherent imaging that allows three-dimensional imaging of complex refractive index (RI). To do so, we show that coherent SI is mathematically equivalent to a superposition of angled illuminations. It follows that raw acquisitions for standard SI-enhanced quantitative-phase images can be processed into complex electric-field maps describing sample diffraction under angled illuminations. Standard diffraction tomography (DT) computation can then be used to reconstruct the sample's 3D RI distribution at sub-diffraction resolutions. We demonstrate this concept by using a SI-quantitative-phase imaging system to computationally reconstruct 3D RI distributions of human breast (MCF-7) and colorectal (HT-29) adenocarcinoma cells. Our experimental setup uses a spatial light modulator to generate structured patterns at the sample and collects angle-dependent sample diffraction using a common-path, off-axis interference configuration with no moving components. Furthermore, this technique holds promise for easy pairing with SI fluorescence microscopy, and important future extensions may include multimodal, sub-diffraction resolution, 3D RI and fluorescent visualizations.

# I. Introduction

Refractive index (RI) is an important optical property in biological objects that is often exploited to visualize endogenous biological contrast with microscopies that specialize in label-free imaging. Standard examples of such microscopies that are heavily used in the biological sciences include phase-contrast and differential interference microscopies. Unfortunately, these techniques do not generate distributions that can be quantitatively associated with RI, and thus only allow qualitative biological analysis. More advanced label-free imaging includes the set of quantitative-phase (QP) imaging techniques that generates complex-value electric-field maps that quantitatively encode optical-path-length (OPL), the integral of sample RI through the illumination path, into optical amplitude and phase components. However, such QP techniques also do not directly yield biological RI values. Furthermore, though QP distributions are often topographically visualized, they do not offer true 3D visualizations and can be corrupted by out-of-focus sample features [1, 2].

In response, several optical techniques have emerged that allow 3D, optically-sectioned, biological visualization with quantitative endogenous contrast. Of most relevance to this work is the set of diffraction tomography (DT) techniques that reconstructs 3D RI distributions from a series of diffraction patterns acquired from illuminating the sample with tilted plane waves [3, 4]. Furthermore, due to synthetic aperture principles, this type of DT naturally synthesizes an imaging aperture larger than the physical aperture set by the microscope and thus allows sub-diffraction resolution coherent imaging [5-8]. Due to its capability for 3D RI visualization, DT has been exploited to noninvasively probe for single-cell biophysical and biochemical properties, and has been used in studies of whole-cell spectroscopy, dry-mass, and morphology [8-10].

Unfortunately, RI has no inherent molecular specificity, which makes it an unsuitable source of contrast for a host of biological studies that require molecular-specific visualization, such as studies of protein localization, gene expression, organelle dynamics, intercellular transport, etc [11-13]. For these studies, fluorescence remains the dominant choice for imaging contrast, and allows biological analysis complementary to that allowed by RI imaging. Thus, to study important biological questions that have significant biophysical, biochemical, and molecular components, multimodal RI and fluorescence imaging may be necessary. Unfortunately, confocal, light-sheet, and multi-photon microscopies, which are the current standards for 3D fluorescent imaging, have system designs that are drastically different from those of DT techniques and hence do not lend themselves to easy incorporation with DT for 3D multimodal imaging at single-cell levels.

We demonstrate here that structured illumination (SI) microscopy, an imaging technique commonly associated with fluorescent super-resolution and a widefield technique that can also enable optical-sectioning for 3D fluorescent imaging [14-16], can also allow 3D reconstruction of RI distributions if operated in the coherent imaging realm. Our technique builds on our previous work for SI-enhanced QP imaging but reformulates SI's coherent imaging process into a multiplexed version of tilted plane-wave illumination [17, 18]. We start by collecting a sequence of raw acquisitions consisting of sinusoidal sample illuminations with varying spatial frequencies, rotations, and translations, interfered with an off-axis reference wave to extract complex-value electric-field maps. These extracted electric-fields under SI illumination were then decomposed into a set of electric-fields that would have been measured under single tilted plan-wave illuminations. Standard DT reconstruction was used to combine these computed electric-field maps into a 3D RI distribution of the sample. Though we focus on using SI for 3D RI visualization in this work, we emphasize that SI is compatible with both 3D RI and fluorescent imaging, and is thus a promising candidate technique for future 3D multimodal applications.

## II. Framework for SI-enabled DT

We begin by developing the framework for 3D RI reconstruction within the regime of coherent SI imaging. We start by describing the intuition behind the concept.

### A. Plane-wave decomposition for coherent SI

In an optical system that images via coherent diffraction, the electric-field at the image plane is linearly related to the electric field at the object-plane via convolution with the system's coherent point-spread-function (PSF) [19]. Hence, Fourier theory directly shows that the image plane's electric-field frequency spectrum is equivalent to the sample plane's electric-field spectrum after being low-passed by the system's coherent transfer function (CTF). Diffraction theory shows this CTF to be a spherical shell, with a circular top-hat projection and a circular arc cross-section in lateral and axial frequency space, respectively [20].

When a diffractive object is illuminated by an orthogonal plane wave with a flat phase-front, the object-plane's electric-field is simply given by the object's amplitude transmittance function (ATF). Hence, the DC-centered region of the object 's diffraction-spectrum passes through the system's TF and the formed image will simply be a DC-centered low-passed version of the object's ATF, as illustrated in Fig. 1(a) below. For tilted plane-wave illumination, the electric-field at the object-plane is a product of the object's ATF with the angled illumination wave-front, mathematically described as a phase-ramp. From Fourier theory, this results in an angular tilt of the sample's diffraction into the system's aperture and a frequency-shift (equivalent to the illumination tilt angle) of the object's diffraction-spectrum through the system's CTF. Thus, the electric-field at the image plane will contain spatial frequencies from a DC-offset region of the object's frequency-spectrum, as shown in Fig. 1(b,c). It follows that SI, which is achieved by interference of tilted coherent plane-waves through the object, multiplexes regions of the object's frequency spectrum through the imaging aperture, with each region corresponding to the angular tilt of an individual wave (Fig. 1(d)). The tilt angles of these plane-waves are directly related to the spatial-frequencies present in the SI pattern.

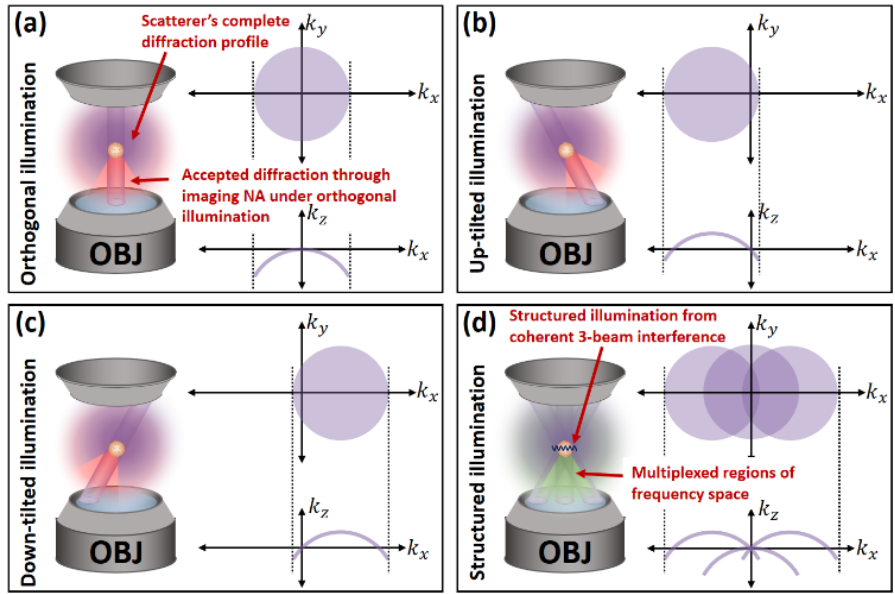

**Fig. 1.** Illumination/detection schemes and corresponding 3D regions of object spatial frequencies imaged at the detector are shown for **(a)** orthogonal, **(b,c)** tilted, and **(d)** structured illumination. Note that the subsection of the object's diffraction that passes through the system aperture under orthogonal illumination (shown in red in (a)) tilts based on illumination angle.

## B. Coherent SI to fill out 3D frequency space

The 3D TF for coherent imaging has been well established by diffraction theory to be a spherical cap mathematically described as a subsection of Ewald's sphere. The radius and subtend angle are given by the imaging system's wavenumber and detection numerical aperture (NA). Because this TF has no axial frequency support, a coherent imaging system operating with typical plane-wave illumination (e.g. standard QP microscopes) offers limited axial sectioning capabilities. DT techniques can circumvent this issue by sequentially illuminating a 3D object at various angles to shift regions of the object's frequency spectrum through the TF. The total frequency content probed by the end of a DT acquisition sequence lie on spherical shells in the object's frequency spectrum, each of which trace Ewald's sphere for each individual illumination wavevector. If a sufficient number of finely incremented illumination angles are used, the object's 3D frequency spectrum can be effectively filled, up to the limit allowed by the system's finite NA. In the case of matched illumination and detection NA, this filled region will form a torus-like shape in 3D frequency space with twice the lateral diffraction limit [7].

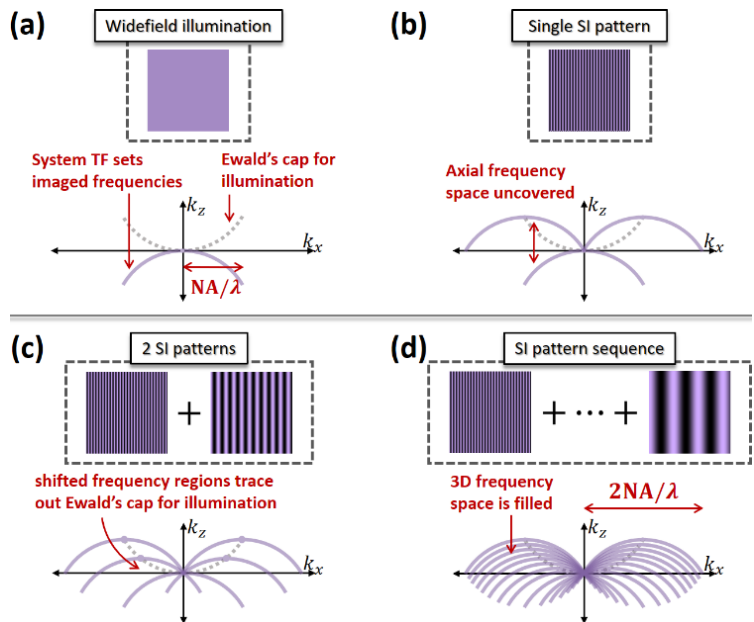

**Fig. 2.** Illustrating concept of SI-enabled DT. **(a)** Widefield coherent imaging samples a spherical shell of object's spectrum. **(b)** SI allows sampling of pairs of shells from regions of the object's spectrum. **(c)** As spatial frequency of the SI changes, the regions of imaged content trace Ewald's sphere of possible illumination wavevectors. **(d)** With sufficiently small increments of SI spatial frequency, 3D region of object's frequency space is filled.

Typical DT methods directly scan the illumination angle via a pair of scan mirrors. Following the principles introduced in the previous section, SI can accomplish the same effect by simply imaging sinusoidal patterns onto the object at incremented spatial frequencies. We illustrate this concept in Fig. 2 below. For each spatial frequency, the SI pattern is translated to analytically separate the multiplexed regions of the object's frequency spectrum, in a manner reminiscent of fluorescent SI. Note that simply illuminating with the highest-allowed spatial frequency, as is done in fluorescent SI to maximize resolution gain, is not sufficient to fill out 3D frequency space in coherent imaging (Fig. 2(b)).

## C. Extracting multiplexed plane-wave components

We start here by introducing the mathematical framework that describes how to extract the multiplexed components from coherent SI, (which follows closely from previous work [17, 18]). We use off-axis holography to obtain 2D complex-valued electric-field measurements from the image-plane. As shown in Fig. 3(a), a raw hologram acquired with sinusoidal SI (experimental methods described in following section) exhibits modulations from both the off-axis interference and the SI pattern. The Fourier transform of the interferogram (Fig. 3(b)) shows the central ambiguity term, the electric-field spectrum at the image-plane, and its conjugate. The image-plane spectrum (outlined in yellow) is digitally filtered out and reset to the center of Fourier space (Fig. 3(c)). The inverse Fourier transform of this gives the electric-field at the image-plane, which can be mathematically expressed as:

$$y(\boldsymbol{r_T}) = h(\boldsymbol{r_T}) \otimes \left[ x(\boldsymbol{r_T}) \cdot \left( h(\boldsymbol{r_T}) \otimes i(\boldsymbol{r_T}) \right) \right] \tag{1}$$

where $\boldsymbol{r_T} = (x, y)$ is the 2D spatial vector, $y(\boldsymbol{r_T})$ is the electric-field at the camera, $x(\boldsymbol{r_T})$ is the sample's 2D complex transmittance function, $i(\boldsymbol{r_T})$ is the electric-field incident at the object plane due to the illumination, $h(\boldsymbol{r_T})$ is the system's coherent 2D PSF, and $\otimes$ is the convolution operator. We assume here that the system's illumination and detection arms use equivalent NA and that magnification is neglected. We express all variables in this section in the 2D domain to remain consistent with standard QP imaging techniques that approximate the electric-field incident at the image-plane to a projection of the object's 3D RI distribution along the illumination path. In the frequency domain, Eq. (1) transforms to:

$$Y(\boldsymbol{k_T}) = H(\boldsymbol{k_T}) \cdot \left[ X(\boldsymbol{k_T}) \otimes \left( H(\boldsymbol{k_T}) \cdot I(\boldsymbol{k_T}) \right) \right] \tag{2}$$

where $\boldsymbol{k_T} = (k_x, k_y)$ is the 2D spatial-frequency vector, $Y(\boldsymbol{k_T}), X(\boldsymbol{k_T}), I(\boldsymbol{k_T})$, and $H(\boldsymbol{k_T})$ are the Fourier transforms of $y(\boldsymbol{r_T}), x(\boldsymbol{r_T}), i(\boldsymbol{r_T})$, and $h(\boldsymbol{r_T})$, respectively. Here, $H(\boldsymbol{k_T})$ is the mathematical projection of the 3D system TF into 2D lateral-frequency space, and is often used as a 2D TF for coherent imaging. We set $H(k_T)$ to reject all wavevectors with lateral-frequencies of magnitudes greater than $k_c$, and accept all others.

In the case of sinusoidal illumination by 3-wave interference, we write the illumination field as: $i(r_T) = 1 + m\cos(k_{c,T} \cdot r_T + \phi_n)$, where $m \leq 1$ and $|k_{c,T}| \leq k_c$. After Fourier transforming and substituting into Eq. (2), we see that the electric-field at the image plane after SI is given by:

$$Y_{SI}(k_T) = H(k_T) \cdot [X(k_T) + m/2\, X(k_T + k_{c,T})e^{j\phi_n} + m/2\, X(k_T - k_{c,T})e^{-j\phi_n}] \quad (3)$$

We see here that the electric-field at the image-plane is a multiplex of the DC-centered, up-shifted, and down-shifted object frequency-spectra terms, which can individually be obtained with sequential tilted plane-waves. This is easily verified by Fourier transforming the individual 2D expressions for the three component waves in the illumination, $i_{DC}(r_T) = 1$, $i_+(r_T) = \exp(j\, k_{c,T} \cdot r_T)$, $i_-(r_T) = \exp(-j\, k_{c,T} \cdot r_T)$, and substituting into Eq. (2). Yellow arrows in Fig. 3(c) indicate the strong DC signals from these plane-waves that are multiplexed into $Y_{SI}(k_T)$. Because $Y_{SI}(k_T)$ is a linear summation of the object's frequency-shifted terms, translating the SI pattern to vary $\phi_n$ allows linear solution of the raw component terms $H(k_T)X(k_T)$, $H(k_T)X(k_T + k_{c,T})$, and $H(k_T)X(k_T - k_{c,T})$, where $H(k_T)$ acts as a DC- centered window over shifted regions of the object's spectrum. Correcting for these shifts result in the terms, $H(k_T)X(k_T), H(k_T - k_{c,T})X(k_T),$ and $H(k_T + k_{c,T})X(k_T)$, which effectively summarize the objective of SI to probe different regions of an object's spectrum with diffraction-limited windows. Inverse Fourier-transforming these corrected terms yield the background-subtracted, electric-field QP maps $x_{DC}(r_T), x_+(r_T),$ and $x_-(r_T)$. We show both the raw component terms, and the associated QP maps in Fig. 3(d-f). Note that $x_{DC}(r_T)$ is exactly the output expected from standard QP microscopy with orthogonal illumination.

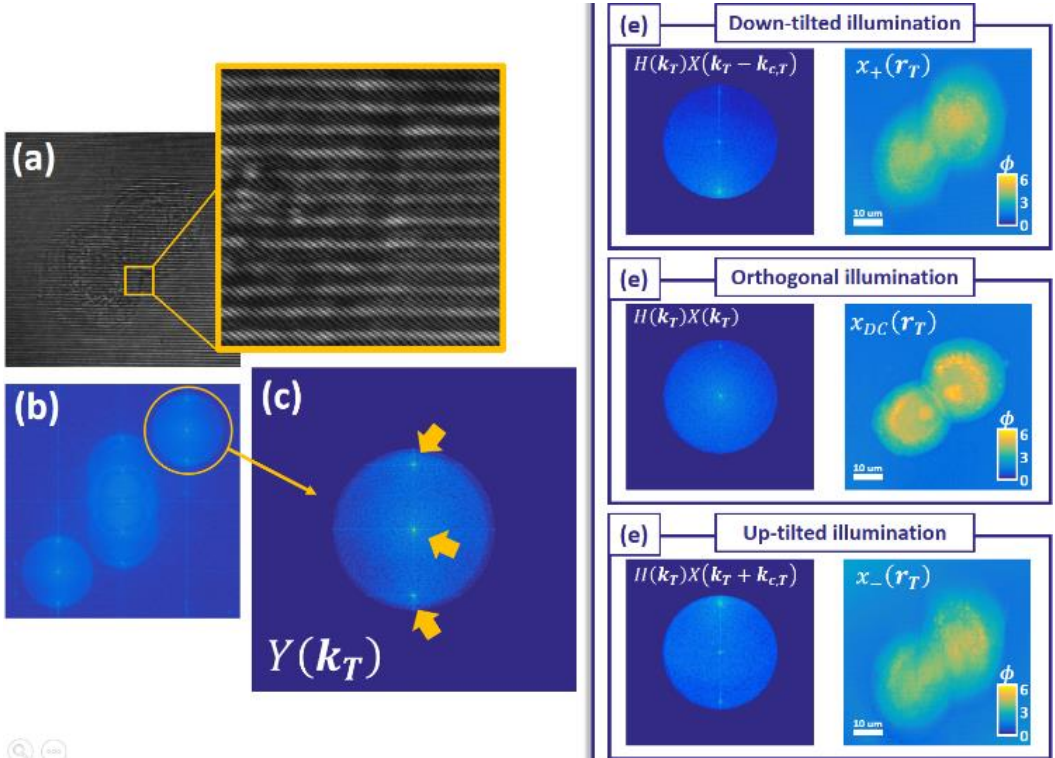

**Fig. 3.** (a) Raw interferogram with SI and (b) associated frequency spectrum (amplitude) are shown. (c) Image-plane frequency spectrum is digitally filtered. (d-f) Frequency spectra of individual plane-wave components, corresponding to down-shifted, up-shifted, and DC-centered regions of the object's frequency spectrum, are analytically solved and shown alongside associated spatial electric-field QP maps.

## D. Tomographic reconstruction of 3D RI

The mathematical framework discussed in this section largely adapts the DT framework introduced in [21] and [22] to reconstruct 3D RI distributions from 2D measurements taken with coherent SI. We strive to keep similar notation to allow easy comparison with Sung et al. [3] specifically. We first introduce the scattering potential, which gives information about the object's RI:

$$f(r) = -k_\lambda^2 (n(r)^2 - n_m^2) \tag{4}$$

where $r = (x, y, z)$ is the 3D spatial vector, $k_\lambda = 2\pi/\lambda$ is the free-space wavenumber, $n(r)$ is the object's 3D RI distribution, and $n_m$ is the constant media RI. The scattering potential is formulated directly from the wave equation under the implicit assumption that the object is semi-transparent, such that the total diffracted wave after passing through the sample $u(r)$ is a sum of the incident, $u_i(r)$, and sample-specific-scatter, $u_s(r)$, waves, respectively:

$$u(r) = u_i(r) + u_s(r) \tag{5}$$

Approximating the object as a lightly-scattering sample, where $u_i(r) \gg u_s(r)$, we can express the sample-specific scatter wave as:

$$u_s(r) = -\int g(|r - r'|) f(r) u_i(r) \, dr \tag{6}$$

where $g(|r - r'|) = \exp(j\, n_m k_\lambda |r - r'|) / (4\pi |r - r'|)$ is the Green's function describing an attenuating spherical wave arising from a point-source at $r = r'$. Plane-wave decomposition of the Green's function and substitution back into Eq. (6) yields the Fourier Diffraction Theorem, which shows that the scattered-wave at the image-plane ($z = 0$) is related to the scattering potential:

$$F(k - k_0) = \frac{jk_z}{\pi} U_s(k_T; z = 0) \tag{7}$$

where $k_0$ is the 3D illumination wavevector, $F(k)$ and $U_s(k_T; z = 0)$ are the 3D and 2D Fourier transforms of $f(r)$ and $u_s(r; z = 0)$, respectively, and $k = (k_T, k_z)$ is the 3D spatial-frequency vector under the constraint $k_z = \sqrt{(n_m k_\lambda)^2 - |k_T|^2}$. Furthermore, due to quasi-monochromatic illumination, which sets the additional constraint $|\mathbf{k_0}| = n_m k_\lambda$, the 3D illumination wavevector is completely determined by its 2D counterpart and is given by $\mathbf{k_0} = (\mathbf{k_{0,T}}, k_{0,z})$, where $k_{0,z} = \sqrt{(n_m k_\lambda)^2 - |\mathbf{k_{0,T}}|^2}$. As is clear here, $U_s(\mathbf{k_T}; z = 0)$ directly relates to the values of $F(\mathbf{k})$ taken along a spherical surface (i.e., the Ewald sphere) displaced by $\mathbf{k_0}$. We can easily correct for this displacement by rewriting E. (7) as:

$$F(\mathbf{k}) = \frac{j(k_z + k_{0,z})}{\pi} U_s(\mathbf{k_T} + \mathbf{k_{0,T}}; z = 0) \tag{8}$$

Sung et al. [3] used a form similar to this for 3D RI reconstruction, where they used the Born and Rytov approximations to relate $U_s(\mathbf{k_T} + \mathbf{k_{0,T}}; z = 0)$ to electric-field QP maps measured from angled plane-wave illuminations, and concluded that the Rytov approximation displayed superior DT reconstruction performance. This conclusion has been reaffirmed by other works [22-25], and we now describe how we applied the Rytov approximation to data generated by SI.

As shown in the previous section, SI enables decoupling of individual, background-subtracted, electric-field maps, $x_{DC}(\mathbf{r_T})$, $x_+(\mathbf{r_T})$, and $x_-(\mathbf{r_T})$. These maps could be equivalently obtained by sequential illumination with the plane-wave components forming the SI, say $i_{DC}(\mathbf{r}) = \exp(j\,\mathbf{k_{DC}} \cdot \mathbf{r})$, $i_+(\mathbf{r}) = \exp(j\,\mathbf{k_+} \cdot \mathbf{r})$, and $i_-(\mathbf{r}) = \exp(-j\,\mathbf{k_-} \cdot \mathbf{r})$, respectively. In the example presented in the previous section, where a SI pattern with 2D spatial-frequency vector $k_{c,T}$ was used, the wavevectors for these plane-waves are $k_{DC} = (0,0,k_\lambda)$, $k_+ = (k_{c,T}, k_{c,z})$, and $k_- = (-k_{c,T}, k_{c,z})$, where $k_z = \sqrt{(n_m k_\lambda)^2 - |k_{c,T}|^2}$. Thus, each background-subtracted electric-field reconstruction, which we now generally refer to as $x_m(r_T)$, can be associated with an effective illumination wavevector, say $k_m$. For a complete SI dataset with acquisitions taken with sinusoidal SI patterns undergoing $s$ rotations with $t$ increments in spatial-frequency per rotation, we build a collection of electric-field maps with associated illumination wavevectors, $\{x_m(r_T), k_m \mid m = 1, 2, \ldots 3 \cdot s \cdot t\}$.

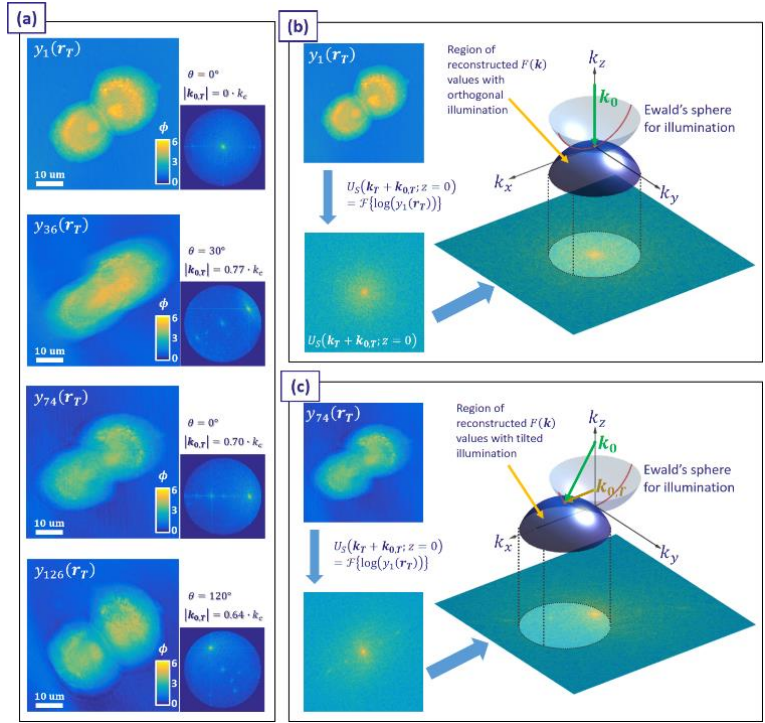

**Fig. 4.** (a) Examples of reconstructed electric-fields and associated Fourier spectra from SI dataset are shown. 2D Rytov-approximated scattered wave is mapped to 3D Ewald surface through $F(\mathbf{k})$ for (b) orthogonal and (c) tilted illumination.

Sung et al. [3] showed that under the Rytov approximation, $U_s(k_T + k_{0,T}; z = 0) = \mathcal{F}\{\log(x_m(r_T))\}$ for each illumination wavevector $k_0 = k_m$, where $\mathcal{F}\{.\}$ and $\log(.)$ are the Fourier and complex logarithm operators, respectively. Thus, for each $1 \leq m \leq 3 \cdot s \cdot t$, values of $F(k)$ along spherical-surfaces in 3D frequency space can be reconstructed by substituting $\mathcal{F}\{\log(x_m(r_T))\}$ and $k_m$ for $U_s(k_T + k_{0,T}; z = 0)$ and $k_0$, respectively, in Eq. (8) above. We illustrate this process in Fig. 4. After $F(k)$ is sufficiently reconstructed, inverse Fourier transform directly gives back the estimate for $f(r)$. 3D RI can then be easily solved using Eq. (4).

# III. Experimental methods and results.

## A. Optical System.

We experimentally demonstrated SI's DT capabilities by using a variant of our original SI-DPM system introduced in [18]. Our current optical system, illustrated in Fig. 5 below, used broadband, single-mode, illumination at $\lambda = 488 \pm 15$ nm (NKT Photonics, EXW-6). The SI pattern was generated by programming a sinusoidal pattern onto an amplitude spatial-light-modulator (SLM, Holeye, HED 6001), which was then imaged through a polarization beam-splitter (PBS, Thorlabs, PBS 251) onto the object through a system of lenses. An adjustable iris diaphragm (F) placed at a Fourier conjugate plane to the SLM spatially filtered out all diffraction orders except the ±1 and 0 orders. These three orders were focused through the condenser objective lens (OBJ, 63X, 1.4 NA Zeiss) to create a 3-beam sinusoidal interference onto the sample. The diffraction from the sample was imaged in transmission through another objective lens (OBJ, 63X, 1.4 NA Zeiss) into a conventional diffraction-phase setup. A ronchi grating (RG, Edmund Optics, Ronchi 70 lpmm) split the signal into three main diffraction orders. A mask, which included a 20 um pinhole (PH, Edmund Optics, 52-869), was used to spatially filter the 0th order to create a common-path, phase-stable, off-axis reference beam to interfere with the 1st order and create an off-axis hologram at the camera plane. The -1st order (and all other extraneous orders arising from RG) was blocked by the mask. We note that, in the case of general widefield imaging (when all pixels on the SLM are turned on), the 0th and ±1st orders arising from RG each contained only one spatial-frequency component. In the case of SI, however, the 0th and ±1st order from the RG each individually contained the spatial-frequency components present the SI pattern, i.e., the 0th and ±1st orders from the sinusoidal pattern written onto the SLM. Due to the spectrally broadband nature of the illumination, each spatial-frequency component in the plane of PH was spectrally dispersed with respect to its order number in relation to the RG and SLM. However, the central 0th SLM/RG-order component represents the purely transmitted portion of the illumination that remained undispersed and physically static regardless of the rotation or spatial frequency magnitude of the SI pattern written onto the SLM. Thus, this component was suitable for spatial filtering by PH (Fig. 5(c)).

Our diffraction-limited lateral coherent resolution (defined as the minimum resolvable spatial period) for typical QP imaging is expected to be $\lambda/\text{NA} = 350$ nm. Due to identical condenser and detection lenses, the SI patterns are also diffraction-limited. Thus, final RI reconstructions are expected to have lateral resolution double that of typical QP, and be able to resolve spatial frequencies of period $\lambda/2\text{NA} = 180$ nm. The resolvable period of axial spatial frequencies is expected to be $\lambda/n(1-\cos\theta) = 520$ nm, where $n = 1.51$ and $\theta$ are the immersion index and maximum accepted half-angle of light, such that $\text{NA} = n\sin\theta$. To generate a 3D RI distribution, 960 raw holograms were acquired (SI patterns underwent 5 translations per spatial-frequency, 32 spatial-frequencies per rotation, and 6 rotations; see Visualization 1) with 15 ms per acquisition. Periodic reconstruction artifacts associated with SI reconstruction were notch-filtered out in frequency space [26].

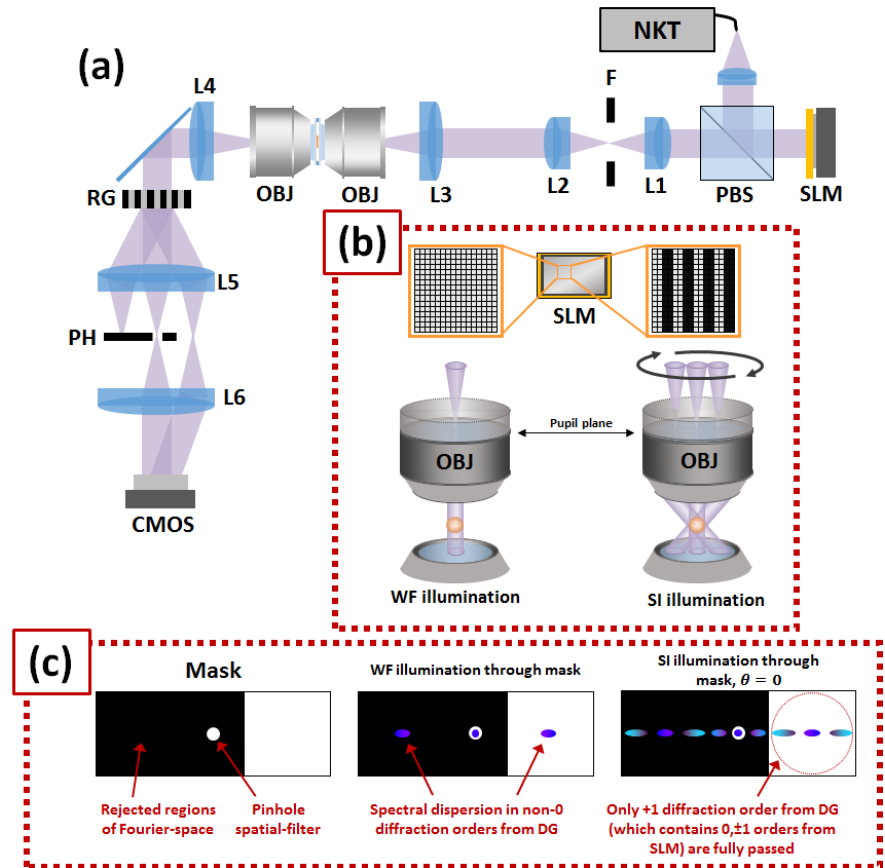

**Fig. 5.** (a) Optical schematic incorporating an SLM into an SI-DPM setup. (b) Periodic patterns programmed into the SLM result in multiple-beam interference at the sample. (c) The mask in the plane of PH achieves common-path, off-axis interference with broadband illumination by spatially filtering the undispersed central component of the 0$^{th}$ diffraction order from RG, regardless of WF or SI illumination.

## B. QP vs RI visualization of polystyrene microspheres

To verify SI's capabilities to quantitatively reconstruct 3D RI distributions, we imaged a sample consisting of a monolayer of 7.3 um polystyrene microspheres immersed in oil (with calibrated $n_m(\lambda) = 1.594$ at $\lambda = 488$ nm). Figs 6(a,b) show the in-focus lateral visualization of a select sample region after QP and RI reconstruction, respectively. Figs 6(c,d) compare the 3D optical-sectioning capabilities between the two reconstruction schemes by taking an axial cross-cut from the region indicated by the dashed-white line in Figs 6(a,b), respectively (3D QP signal was reconstructed by digitally propagating the in-focus signal in Fig. 6a using Fresnel kernels). As is evident, RI reconstruction demonstrates clear optical sectioning by depth- localizing RI signal from the microsphere. Conversely, the QP signal shows no such depth-localization. We note that though Fig. 6(b) demonstrates a clear circular RI cross-section laterally, Fig. 6(d) shows that the axial RI cross-cut is elongated axially. This deformation is due to the "missing-cone" problem, which results from limited axial frequency-space coverage for the lower lateral frequencies. This issue is inherent to beam-scanning DT techniques and is typically addressed with computational methods [27].

We now quantitatively evaluate the reconstructed QP and RI values. Fig. 6(e) shows the QP profile measured across the microsphere indicated by the dashed-white line in Fig. 6(a), and shows a QP peak at ~0.6 rad. This profile matches the theoretical QP profile expected from a microsphere with RI of $n_{PS} = 1.601$. This measured RI value falls to within 1% of the theoretical RI value for polystyrene $\widehat{n_{PS}}(\lambda) = 1.605$ for $\lambda = 488$ nm, based from previous work fitting spectral RI measurements with the Sellmeier dispersion formula [28].

Unlike the QP signal, which integrated RI along OPL and demonstrated an elliptical profile in Fig. 6(e) for a spherical particle with homogenous RI, reconstructed RI values are localized to specific points in 3D space. Thus, the RI profile across a cross-section of a microsphere is expected to take the form of a rectangular function bound between the RI of the polystyrene microsphere and the surrounding media. Fig. 6(f) shows the measured RI profile across the same region measured in Fig. 6(e) and clearly demonstrates a rectangular profile bound between $n_{PS}$ and $n_m$.

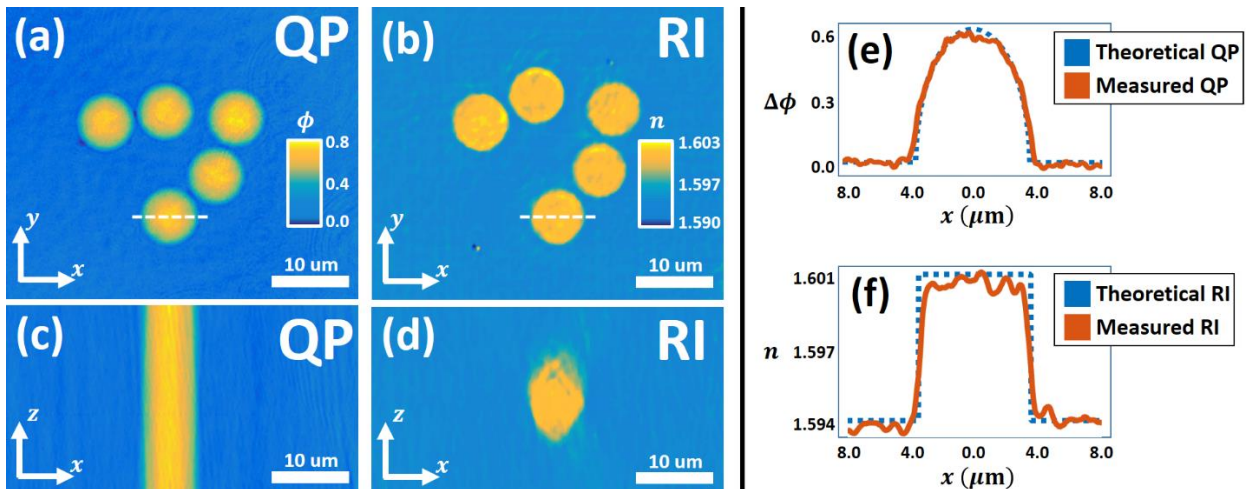

**Fig. 6.** (a,b) QP and RI lateral visualizations, respectively, of 7.3 um polystyrene microspheres. (c,d) QP and RI axial cross-cuts, respectively, across a single microsphere. (e,f) Quantitative profiles are shown comparing measured QP and RI values to theoretical.

## C. 3D RI reconstruction of human breast adenocarcinoma

We first demonstrate 3D biological RI reconstruction of cells from the human breast adenocarcinoma (MCF-7) line. These cells are popularly used in studies of tumor biology and previous works have extensively studied their signaling pathways, gene-activation events, molecular receptors, hormonal responsiveness, and proliferation rates [29]. MCF-7 cells were cultured using Dulbecco's Modified Eagle Medium (DMEM) supplemented with 10% fetal bovine serum, 10 µg/mL and 1µL/mL pen-strep. Cells were plated at low density onto #1.5 coverslips and allowed to attach overnight. Cells were subsequently fixed using 4% formaldehyde in PBS for 10 minutes, and were then washed and incubated with PBS prior to imaging.

In Fig. 7, we compare 3D RI reconstruction with typical QP visualization of a single MCF-7 cell at axial slices spaced 2.26 µm through the cell volume. Standard Fresnel propagation kernels were used to digitally propagate the 2D QP map. We see immediately from both RI (Fig. 7(a-d)) and QP (Fig. 7(e-h)) visualizations that the specific imaged cell has a high-mass bulk center-region with low-mass surrounding cytoplasmic extensions. However, as seen in Fig. 7(e-h), QP does not allow high contrast visualization of structures within the high-mass region, and in fact does not allow the viewer to even make an educated guess about whether the region is simply "thick", which relates to cell morphology, or "dense", which relates to cell composition. This is of course due to RI and physical-path-length being entangled into OPL, the source of contrast in QP imaging. Furthermore, the various axial slices show almost no variation in QP signal, as shown in Fig. 7(e), which indicates two cytoplasmic extensions that are visualized even after axial defocus. This conforms to our expectation that QP imaging has little axial resolution. In contrast, the RI visualizations of the same axial slices show a drastic increase in intra-cellular contrast, which can be attributed to both optical sectioning as well as the doubled lateral resolution over QP imaging. An obvious advantage of RI imaging includes significant enhancement of image contrast when visualizing the central high-mass region of the cell. Unlike the ambiguities associated with QP imaging, RI shows that this region has a higher "density" (corresponding to RI values of ~1.335) than the surrounding cytoplasm, which suggests a compositional change. In fact, Fig. 7(c) shows that this region has a circular delineation and is markedly denser than the surrounding cytoplasm, which suggests that this region is dominated by one large sub-cellular component. After referencing previous works that show molecular-specific fluorescent labelling of MCF-7 cells, and drawing comparisons to the size, shape, and positioning of the high-density region, it is reasonable to hypothesize that this high-density region is the cell's nucleus. Even within the nucleus, high density structures can be clearly visualized. Given the shapes and locations of these structures, we hypothesize that they are the nucleolus and endoplasmic- reticulum. The nucleolus shows a grainy texture in Fig. 7(c), which conforms with previous works visualizing MCF-7 nucleoli via fluorescence [30]. The surrounding cytoplasm also exhibits significantly increased contrast, and various cytoplasmic extensions and cavities are clearly visible in sharp focus in Fig. 7(a-c). These cytoplasm structures were either not visible or out-of-focus with QP imaging (Fig. 7(e-h)). Furthermore, RI visualization shows both the nucleus and cytoplasm changing shape as the imaged depth-plane is axially moved, which demonstrates 3D optical sectioning (see Visualization 2).

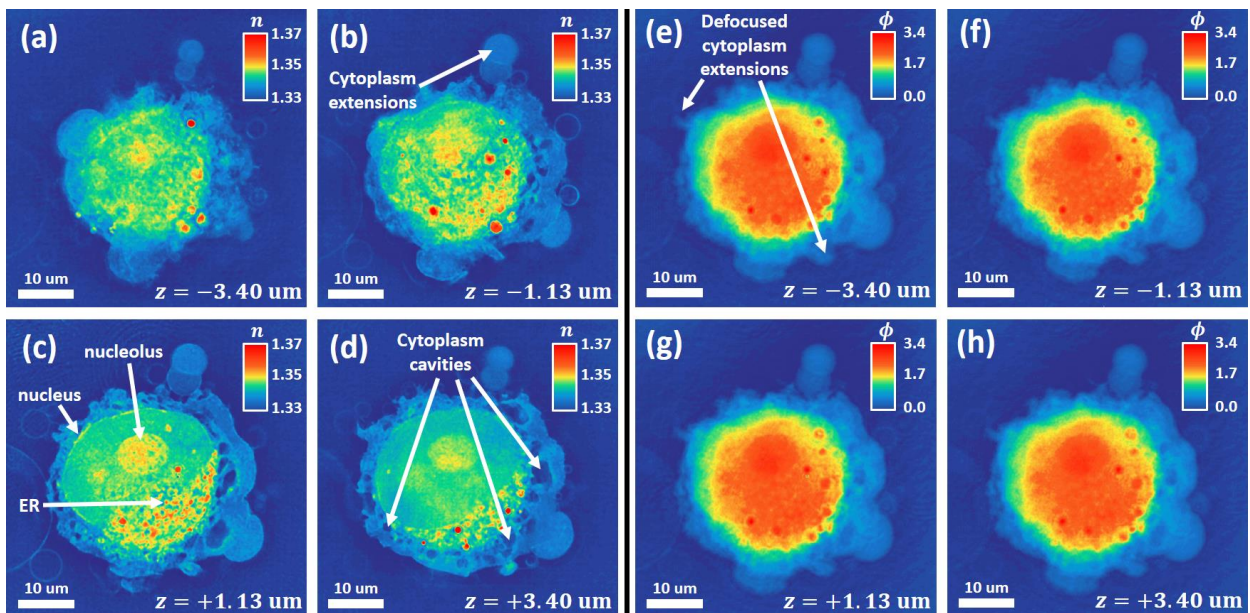

**Fig. 7.** Axial slices are shown from an MCF-7 cell after (a-d) 3D RI reconstruction via SI and (e-h) Fresnel propagating a QP image. RI visualization shows clear image contrast, resolution and optical-sectioning enhancement over QP.

Interestingly, we have observed (not shown here) MCF-7 cells from the same population that exhibit nuclear RI values lower than those of those of the rest of the cell. Furthermore, most MCF-7 cells do not have as high of a karyoplasmic ratio (nucleus/whole-cell volume ratio). Indeed, one of the fundamental properties of eukaryotic cells is an ability to maintain nuclear sizes in relation to whole-cell volume, which is crucial for many nuclear-specific cell functions, such as nuclear transport, gene expression, organization of inter-nuclear components and compartments (nucleolus, Cajal bodies, Kremer bodies, etc), and general cell regulation [31]. Given the expected consistency of nuclear size, nuclear growth is typically associated with a progression through the cell-cycle [32]. We hypothesize that the enlarged nucleus in Fig. 7 indicates this individual cell's preparation, before fixation, to enter into cell division.

## D. 3D RI reconstruction of human colorectal adenocarcinoma

We now demonstrate 3D biological RI reconstruction of human colorectal adenocarcinoma (HT-29) cells. Like the MCF-7 cells presented in the previous section, the HT-29 cell line is widely used to study tumor biology. However, HT-29 cells have a drastically different morphology than MCF-7, and typically exhibit globular shapes. HT-29 cells have been popular imaging choices in previous DT works [3, 33], perhaps because of their rounded, roughly spherical, profiles, and so we also image them to further explore 3D RI reconstruction via SI. The HT-29 cells were grown in McCoy's 5A Modified Medium supplemented with 10% fetal bovine serum and 1 uL/mL pen-strep antibiotic. Subsequent fixation steps with formaldehyde were performed in a similar fashion as to the MCF-7 cells from above.

In Fig. 8, we demonstrate 3D RI reconstruction of a pair of conjoined HT-29 cells, and visualize their hallmark globular structure (see Visualization 3 for comparison between RI and QP 3D reconstruction). In Fig. 8(a-d), we show axial slices taken through the RI tomogram taken at 3.8 µm increments. Due to the cells' globe-like profiles, clear differences in the axial slices are apparent. The cell boundary separating the cell cytoplasm from the surrounding PBS immersion media is clearly visible in all slices and shows the cells' circular cross-sections enlarging and then shrinking as the visualized cross-section shifts axially up through the cell, as is expected when visualizing a globular structure. Fig. 8(a) shows an axial slice at z = -5.67 µm, which contains a number of high-density localizations below the nucleus. Molecular labelling is necessary to truly identify these localizations. However, given their sizes and positions around the nucleus, we hypothesize that these localizations are lipid-based vesicles, which are known to have high RI lipid bilayer. Previous works have demonstrated fluorescent imaging of vesicles and have reported vesicles to congregate adjacent to the nuclear region of the cell [34]. These findings conform to our observation of the vesicles being situated almost directly beneath the nucleus. The next axial slice at z = -1.89 µm is closer to the center of the cell and exhibits larger HT-29 cross-sectional area. Here, we can see a ring-shaped region of high RI that directly surround the nuclear region.

We also see the tail ends of the nucleoli start to appear at this axial position. Interestingly, unlike our observations with the MCF-7 cell shown in Fig. 7, even though the nuclei in the two HT-29 cells show regions with modestly high RI, the nuclei as whole organelles do not exhibit a distinct difference in RI compared to the rest of the cell. Previous studies performing 3D RI tomography on HT-29 cells have demonstrated similar observations [3]. This observation is continued into the next axial slice, at z = +1.89 µm, where now the nucleoli appear fully formed. In Fig. 8(e-g), we show 3D visualization of the RI tomogram after applying RI-thresholded opacity constraints so as to emphasize different biological features. Fig. 8(e) emphasizes the cell boundary that encapsulates the cell. The cells' subcellular components can be viewed through the boundary so as to see the visualize their positioning in the cell's body. Fig. 8(f) emphasizes the intracellular content within the cytoplasm by setting the RI values associated with the cytoplasm (RI < 1.337) transparent. We can see that the moderately-high, intra-cell components, typically with RI of ~ 1.35, to be concentrated in areas around the nuclear region. Fig. 8(g) shows the positioning of high RI components (RI>1.37) to be localized to either the nucleoli or nuclei periphery. Rotational view of the tomogram is shown in Visualization 4.

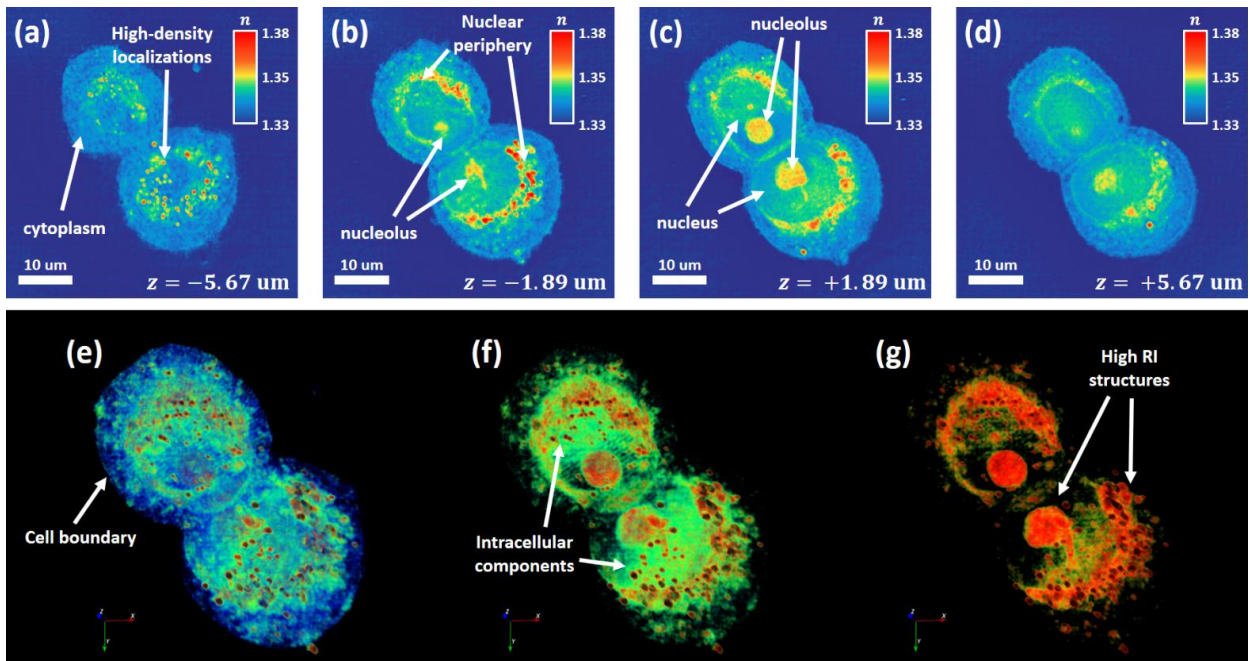

**Fig. 8.** Axial slices are shown from two conjoined HT-29 cells after (a-d) 3D RI reconstruction via SI. The globular structures of the HT-29 cells are clearly visualized with the cells' circular cross-section enlarging and then shrinking as the focus plane is translated upwards. Furthermore, distinct intracellular morphologies are 3D resolved via endogenous RI contrast. Of note are the cell's nucleoli and nuclear periphery, which demonstrate high RI values. (e-f) Tomograms show 3D false-colored RI distributions visualized with different RI-thresholded opacity constraints to emphasize different intracellular features. 3D visualizations were done using Icy, a freely available platform for biological image analysis [35].

# IV. Discussion and conclusion

In this work, we have presented a mathematical framework that draws parallels between coherent SI and DT by first portraying SI as a multiplexed version of tilted-plane wave illumination. We emphasize that the electric-field measured at the image plane after SI of the object is a superposition of the electric-fields that would have been measured after object illumination with the tilted plane-waves individually composing the SI pattern. It follows that SI of the object at various rotations and spatial-frequencies is sufficient to fill out 3D frequency-space, as is required in conventional DT. However, unlike conventional DT techniques which utilize scan mirrors that inherently associate with mechanical instability and illumination-angle ambiguity, SI enables phase-stable electric-field acquisitions with a common-path, off-axis setup with no mechanically moving components, which can thus potentially measure 3D RI with high stability. Furthermore, the common-path off-axis interference configuration also allows spectrally broadband illumination to be used, which improves imaging artifacts from coherent noise. Our optical system used for this work generated SI patterns using an SLM based on nematic liquid crystal technology, which associates with slow response times. We fully expect future improvements of this system to incorporate ferroelectric-based SLMs or digital-micromirror-devices (DMD) for significantly faster generation of SI patterns.

Furthermore, we also note that our theoretical treatment for SI-enabled DT assumed monochromatic illumination. In reality, however, we used illumination with ~30 nm bandwidth, which was previously demonstrated to reduce coherent noise in 2D QP imaging [18]. More notably for 3D RI imaging, broadband illumination also results in imaging 3D frequency space through an Ewald shell of non-uniform, non-infinitesimal axial thickness [33]. It follows that simply broadening the illumination spectrum may allow improvements in axial resolution that could potentially be utilized in conjunction with tilted illuminations [36]. Future extensions of this work may explore optimization of illumination bandwidth with number of illumination tilt angles to efficiently fill 3D frequency space.

Generally, however, this work demonstrates that SI, which is popularly associated with fluorescent super-resolution imaging, also has significant utility for coherent biological imaging applications. Indeed, coherent imaging is uniquely capable of imaging biological objects with high, endogenous contrast, and has been used to extract biophysical and biochemical properties. Applications of this have included quantitative measurements of whole-cell morphology and mass, spectroscopy, hemoglobin concentration, etc [8-10]. Previous work that considered SI's capabilities to augment the effectiveness and accuracy of these applications has utilized SI for sub-diffraction resolution imaging of coherent scatter and optical phase-delays [17, 37-40]. To the best of our knowledge, this work reports one of the earliest examples of applying the SI framework for extracting 3D biological RI, the intrinsic optical property that determines the object's scattering and phase-delay characteristics [41].

Furthermore, this method to extract 3D biological RI lends itself easily to compatibility with SI-based fluorescent super-resolution. Fluorescence imaging is the dominant choice for molecular-specific biological visualization, and is the standard imaging technique used in a host of works studying gene expression, protein interaction, organelle structure, subcellular transport, and general intracellular dynamics [11-13, 42, 43]. As such, fluorescent imaging offers imaging capabilities that directly complement those offered by coherent imaging, which offer no intrinsic specificity. Fluorescent SI has demonstrated that 3D fluorescent resolution-doubling further augments such imaging capabilities, and has been used for visualization of centrosome and nuclear architecture as well as microtubule and mitochondrial dynamics. [15, 16, 44, 45]

Based on this work, we believe that SI holds promise as a multimodal sub-diffraction technique that can be used for 3D high-resolution fluorescent and RI volume imaging. Such a technique can have important applications in studying biological questions that have significant molecular and biophysical/biochemical components.

**Funding.** National Science Foundation (NSF) (1403905);

**Acknowledgment**. We thank Kyoohyun Kim for useful discussions about diffraction tomography and Brianna Loomis for help in preparation of biological samples.